\documentclass[aps,prl,twocolumn]{revtex4}
\usepackage{graphicx}
\usepackage{amsmath}

\begin{document}

\title{Addressing Individual Atoms in Optical Lattices with Standing-Wave Driving Fields}
\author{Jaeyoon Cho}
\affiliation{Division of Advanced Technology, Korea Research Institute of Standards and Science, Daejeon 305-340, Korea}
\date{\today}
\begin{abstract}
A scheme for addressing individual atoms in one- or two-dimensional optical lattices loaded with one atom per site is proposed. The scheme is based on position-dependent atomic population transfer induced by several standing-wave driving fields. This allows various operations important in quantum information processing, such as manipulation and measurement of any single atom, two-qubit operations between any pair of adjacent atoms, and patterned loading of the lattice with one atom per every $n$th site for arbitrary $n$. The proposed scheme is robust against considerable imperfections and actually within reach of current technology.
\end{abstract}
\maketitle

\newcommand{\ket}[1]{\left|{#1}\right>}
\newcommand{\bra}[1]{\left<{#1}\right|}
\newcommand{\abs}[1]{\left|{#1}\right|}

A system of cold neutral atoms trapped in an optical lattice is expected to be a promising architecture for quantum information processing \cite{jz05}. In particular, by using superfluid to Mott insulator phase transition, a large number of atoms can be loaded into an optical lattice in such a way that each lattice site is filled with one atom \cite{jbc98,gme02,sms04,spp07}. Such a Mott insulator state naturally provides a scalable qubit system, where each atom, i.e., each lattice site, stores a qubit encoded into its two ground hyperfine levels. Two-qubit operations in such a system can be achieved by exploiting cold controlled collisions between atoms \cite{jbc99} or atomic dipole-dipole interaction \cite{bcj99}. Although these operations can not be applied selectively between one pair of atoms, they can be applied collectively between every adjacent atom, thereby generating a highly entangled state called a cluster state \cite{mgw03}. Once a cluster state is generated, arbitrary quantum logic operations can be performed simply by measuring individual atoms in appropriate bases \cite{rb01}. For this purpose, an essential requirement is the ability to manipulate and measure single atoms individually.

Addressing individual atoms in optical lattices is, however, still experimentally challenging. The main reason is that the lattice separation of typical optical lattices is given by half the optical wavelength, while atoms are manipulated by lasers having wavelengths comparable to or larger than the optical wavelength. This situation thus seems to imply that the addressing laser beam has to be focused within a length scale smaller than its wavelength, while it is limited by diffraction. 

There have been several approaches to this problem \cite{sch00,dvm02,ppl03,mad06, sdk04,zrs06, yc00,cdj04,jlb06}. One straightforward approach is to increase the lattice separation, which was employed in most of the experimental attempts \cite{sch00,dvm02,ppl03,mad06}. In this case, however, it is hard to load a large number of atoms into the lattice with one atom per site. Another approach is to apply a position-dependent external field so that individual atoms are distinguished by their different resonance frequencies \cite{sdk04,zrs06}. Otherwise, one would use a {\em pointer} atom that serves as a quantum information carrier \cite{yc00,cdj04}. In any case, the realization imposes significantly more demands on the experimental complexity and precision.

In this paper, a new approach to the problem of addressing individual atoms in optical lattices is proposed. One of the advantages of the present scheme is that it is compatible with any one- or two-dimensional optical lattice setup. For instance, it is compatible with a typical lattice setup using an optical wavelength, and also compatible with a non-square (such as a triangular) optical lattice. Another advantage is that  its experimental setup is quite simple: in addition to the original optical lattice setup, we need just several laser beams tuned to the atomic transition frequencies (with some detunings). Moreover, the scheme is robust against experimental imperfections, and actually all techniques involved in it have been experimentally demonstrated. Remarkably, its applicability is not restricted to the unitary rotation and measurement of a single atom. It can be also used for other important purposes, such as two-qubit operations between an arbitrarily chosen pair of adjacent atoms and the patterned loading of an optical lattice with one atom per every $n$th site for arbitrary $n$.

\begin{figure}
\includegraphics[width=0.22\textwidth]{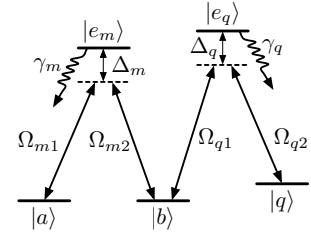}
\caption{The involved atomic levels and transitions. $\Delta_i$, $\gamma_i$, and $\Omega_{ij}$ ($i=\{m,q\}$, $j=\{1,2\}$) denote the detunings, spontaneous decay rates, and Rabi frequencies, respectively.}
\label{fig:level}
\end{figure}

Fig.~\ref{fig:level} depicts the involved atomic levels and transitions. We use three ground levels $\ket{a}$, $\ket{b}$, and $\ket{q}$: among them, $\ket{a}$ and $\ket{b}$ are used to encode a qubit. Raman transitions $\ket{a}\leftrightarrow\ket{b}$ and $\ket{b}\leftrightarrow\ket{q}$ are mediated by excited levels $\ket{e_m}$ and $\ket{e_q}$ with detunings $\Delta_m$ and $\Delta_q$, respectively, where $\gamma_i$ and $\Omega_{ij}$ $(i=\{m,q\}$, $j=\{1,2\})$ denote the corresponding spontaneous decay rates and (complex) Rabi frequencies, respectively. The atomic state at lattice site $s$ will be denoted by a subscript as $\ket{~}_s$.

In order to illustrate the underlying idea, let us first consider the case of a one dimensional optical lattice loaded with one atom per site. We consider the situation where a single atom at lattice site $k$ is manipulated by a well-focused laser beam centered on the atom. We assume that the beam waist is larger than the lattice separation $d_l$, but is still smaller than $2Ld_l$ for a moderately small integer $L$. The manipulating beam thus affects at most $2L+1$ atoms at sites $k-L$ to $k+L$ \cite{jlb06}. The basic idea is, prior to the single-atom manipulation, transferring the atomic states of the neighboring sites coherently into {\em quenching} states that are decoupled from the manipulating beam. As will be shown later, this operation can be performed collectively for all neighboring $2L$ atoms simply by applying a couple of standing-wave fields. We will call this operation a quenching operation \footnote{The term ``quenching'' is used with a similar meaning in the context of nanolithography; See Ref.~\cite{jtd98}. An important difference is that in our case the quenching operation is a coherent process.}. Once the quenching operation has been completed, the manipulating beam will affect only the selected atom at site $k$. For convenience, let us define $S_L^{(k)}$ as the set of lattice sites $s$ such that $s=k+(L+1)n$ for all integers $n$. A quenching operation $Q_L^{(k)}$ is defined such that it does not affect atoms at sites $s\in S_L^{(k)}$ while it transfers the atomic state coherently from the qubit subspace to the quenching subspace as $\ket{a}_s\!\bra{a}+\ket{q}_s\!\bra{b}$ for all atoms at sites $s\not\in S_L^{(k)}$, where state $\ket{q}$ serves as the quenching state. We also define the inverse quenching operation $(Q_L^{(k)})^{-1}$ transferring the state in the opposite direction.

For the moment, let us assume we are able to perform the quenching operation. The applications of it follow rather straightforwardly. Some of them are listed below:

\textit{Single-atom addressing.---}The first one is a single-qubit unitary rotation $R_z^{(k)}(\alpha)=\exp(-i\alpha\sigma_z^{(k)}/2)$, where $\sigma_z^{(k)}=\ket{a}_k\!\bra{a}-\ket{b}_k\!\bra{b}$. In order to perform this operation, we first perform the quenching operation $Q_L^{(k)}$, then apply a focused laser pulse with Rabi frequency $\Omega_{m2}$, and finally perform the inverse quenching operation $(Q_L^{(k)})^{-1}$. If $\Delta_m\gg\abs{\Omega_{m2}}\gg\gamma_m$, the laser pulse induces a unitary rotation $R_z^{(k)}\left(\frac{\abs{\Omega_{m2}}^2}{\Delta_m}\Delta t\right)$, where $\Delta t$ is the pulse duration. Note that this laser pulse has no effect on the neighboring atoms since after performing $Q_L^{(k)}$ there is no population in state $\ket{b}_s$ for $s\not\in S_L^{(k)}$. Another single-qubit rotation about an orthogonal direction $R_x^{(k)}(\beta)=\exp(-i\beta\sigma_x^{(k)}/2)$, where $\sigma_x^{(k)}=\ket{a}_k\!\bra{b}+\ket{b}_k\!\bra{a}$, can be performed with the help of a collective Hadamard operation $H_C=\bigotimes_s\frac{\sigma_x^{(s)}+\sigma_z^{(s)}}{\sqrt2}$, which can be done by applying global pulses addressing every atoms identically. It is easily seen that a sequence of operations $H_C$, $R_z^{(k)}(\beta)$, and $H_C$ results in  the operation $R_x^{(k)}(\beta)$. By combining these two kinds of rotations, an arbitrary single-qubit rotation can be performed. A single-qubit measurement can be also performed in the same fashion: it is done by observing the resonance fluorescence on transition $\ket{b}_k\leftrightarrow\ket{e_m}_k$ induced by a focused beam applied between performing $Q_L^{(k)}$ and $(Q_L^{(k)})^{-1}$.

\textit{Two-qubit operation.---}Another application is a collective CPHASE operation $C_L^{(k)}=\bigotimes_{s\in S_L^{(k)}}(\ket{a}_s\!\bra{a}\otimes I^{(s+1)}+\ket{b}_{s}\!\bra{b}\otimes\sigma_z^{(s+1)})$, where $I^{(s)}$ is the identity operator. This operation also starts from performing the quenching operation $Q_L^{(k)}$. After that, we use the methods outlined in Refs.~\cite{jbc99,bcj99} in such a way that the lattice potential corresponding to state $\ket{b}$ is shifted as much as about the lattice separation and restored back afterwards so that every atoms in state $\ket{b}$ interact with the next neighboring atoms. Note that after performing $Q_L^{(k)}$ only atoms at sites $s\in S_L^{(k)}$ can be in state $\ket{b}_s$. This interaction can be used to perform a conditional operation such that the state acquires a phase $\pi$ only when it is $\ket{b}_s\ket{q}_{s+1}$. By performing $(Q_L^{(k)})^{-1}$ afterwards, the net effect is the same as the desired operation $C_L^{(k)}$. If we use this collective CPHASE operation, even a selective two-qubit operation can be achieved. It is easily seen that a sequence of operations $C_L^{(k)}$, $R_x^{(k+1)}(-\alpha/2)$, and $C_L^{(k)}$ results in a two-qubit controlled rotation $\ket{a}_k\!\bra{a}\otimes I^{(k+1)}+\ket{b}_k\!\bra{b}\otimes R_x^{(k+1)}(\alpha)$ up to a single-qubit rotation $R_x^{(k+1)}(-\alpha/2)$.

\textit{Patterned loading.---}Another straightforward application is the patterned loading of an optical lattice with one atom per every $(L+1)$th site. It is simply done by first preparing all atoms in state $\ket{b}$, then performing $Q_L^{(k)}$, and finally turning off the lattice potential corresponding to state $\ket{q}$.

\begin{figure}
\includegraphics[width=0.3\textwidth]{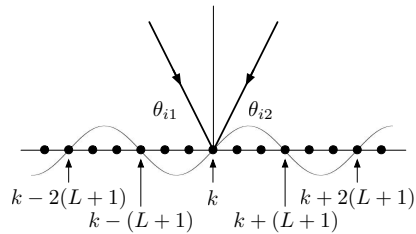}
\caption{Two plane waves incident on the optical lattice tilted respectively by angled $\theta_{i1}$ and $\theta_{i2}$ produce a standing-wave field along the trap axis.}
\label{fig:quenching}
\end{figure}

In the above scenario, the only nontrivial part is how to implement the quenching operation, which we will concentrate on in the remainder of this paper. For a quenching operation, we apply two standing-wave fields which, respectively, induce the transitions $\ket{b}\leftrightarrow\ket{e_q}$ and $\ket{q}\leftrightarrow\ket{e_q}$. Due to the position dependence of the fields, the transitions take place with different Rabi frequencies for each atom. Let us denote the corresponding Rabi frequencies at site $s$ respectively by $\Omega_{q1}^{(s)}$ and $\Omega_{q2}^{(s)}$. The standing-wave field corresponding to $\Omega_{qi}^{(s)}$ ($i=1,2$) is produced by two plane waves having a common wavelength $\lambda_{qi}$ and tilted from the trap axis by angles $\theta_{i1}$ and $\theta_{i2}$, respectively, as depicted in Fig.~\ref{fig:quenching}. If we set $\theta_{i1}=\pm\theta_{i2}=\theta_{qi}$, the spatial period of the standing wave along the trap axis is given by $\frac{\lambda_{qi}}{\cos\theta_{qi}}$, where the position of the nodes is determined by the relative phase between the two plane waves. What we should do is to adjust the involved parameters so that both the standing-wave fields have the same spatial period $\lambda_q=\frac{\lambda_{q1}}{\cos\theta_{q1}}=\frac{\lambda_{q2}}{\cos\theta_{q2}}$ and have the common nodes. Furthermore, we require that each lattice site  in $S_L^{(k)}$ coincides with a node while all other lattice sites never coincide with the nodes. This requirement can be met in multiple ways, but for simplicity we restrict our consideration to the case of
\begin{equation}
\lambda_q/2=(L+1)d_l.
\end{equation}

To be more specific, let us assume the lattice potential is sufficiently deep, thus the tunneling between lattice sites is suppressed. Furthermore, we assume every atoms are in the vibrational ground state. In this case, each potential well can be approximated by a harmonic potential. Let us denote by $a_s$ ($a_s^\dagger$) the annihilation (creation) operator of the vibrational mode at site $s$. The involved evolution of the atomic state is then described, in the rotating frame, by the following Hamiltonian \cite{mlk97}:
\begin{align}
\mathcal{H}_q^{(s)}=&(\Delta_q-i\frac{\gamma_q}{2})\ket{e_q}_s\!\bra{e_q}+\sin(\eta(a_s+a_s^\dagger)+\phi_s)\nonumber\\
&\times[\Omega_{1}(t)\ket{e_q}_s\!\bra{b}+\Omega_{2}(t)\ket{e_q}_s\!\bra{q}+h.c.],\label{eq:hamil}
\end{align}
where $\eta=2\pi x_0/\lambda_q$ is the Lamb-Dicke parameter with $x_0(a_s+a_s^\dagger)$ the position operator of the harmonic oscillation, $\phi_s=\pi(s-k)/(L+1)$, and $\Omega_i(t)$  ($i=1,2$) is the corresponding Rabi frequency at the peak position of the standing-wave field. By using ${}_s\!\bra{0}\sin(\eta(a_s+a_s^\dagger)+\phi_s)\ket{0}_s=e^{-\eta^2/2}\sin(\phi_s)$, where $\ket{0}_s$ denotes the vibrational ground state of the atom at site $s$, the Hamiltonian~(\ref{eq:hamil}) can be written as
\begin{align}
\mathcal{H}_q^{(s)}=&(\Delta_q-i\frac{\gamma_q}{2})\ket{e_q}_s\!\bra{e_q}+e^{-\eta^2/2}\sin(\phi_s)\nonumber\\
&\times[\Omega_1(t)\ket{e_q}_s\!\bra{b}+\Omega_2(t)\ket{e_q}_s\!\bra{q}+h.c.],
\end{align}
which implies the effective Rabi frequencies are given by
\begin{equation}
\Omega_{qi}^{(s)}(t)=e^{-\eta^2/2}\sin(\phi_s)\Omega_i(t).\label{eq:rabi}
\end{equation}
Consequently, we reach an essential condition
\begin{equation}
\Omega_{qi}^{(s)}=0\text{ for }s\in S_L^{(k)},~\Omega_{qi}^{(s)}\neq0\text{ for }s\not\in S_L^{(k)}.
\end{equation}

For $L\le2$, the quenching operation $Q_L^{(k)}$ can be performed straightforwardly by using the Raman transition. For a large detuning $\Delta_q\gg\abs{\Omega_{qi}^{(s)}}$, Raman transition $\ket{b}_s\leftrightarrow\ket{q}_s$ takes place with Rabi frequency $\Omega_{q1}^{(s)}\Omega_{q2}^{(s)}/\Delta_q$. Since this Rabi frequency is the same for all lattice sites $s\not\in S_L^{(k)}$ when $L\le2$, the operation $Q_L^{(k)}$ is achieved simply by applying a Raman $\pi$-pulse. However, this method does not work for $L>2$ since in that case each atom experiences a different Rabi frequency.

\begin{figure}
\includegraphics[width=0.3\textwidth]{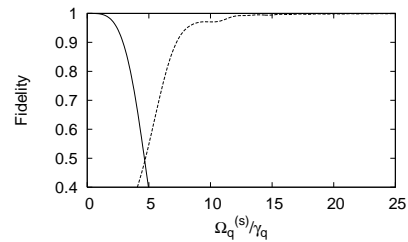}
\caption{Fidelity of state transfer $\ket{\Psi}_s=\frac{1}{\sqrt2}(\ket{a}_s+\ket{b}_s)\rightarrow\ket{\Psi'}_s=\frac{1}{\sqrt2}(\ket{a}_s+\ket{q}_s)$ by STIRAP. The fidelity between the final state and $\ket{\Psi}_s$ ($\ket{\Psi'}_s$) is plotted as a solid (dotted) curve with respect to the peak value $\Omega_q^{(s)}$ of the Gaussian pulses.}
\label{fig:stirap}
\end{figure}

A more powerful way of performing the quenching operation regardless of $L$ is to use the technique of stimulated Raman adiabatic passage (STIRAP) \cite{kts07}. As a typical example of STIRAP, let us consider a case where the standing-wave fields are applied in order as Gaussian pulses
\begin{equation}
\Omega_i(t)=(-1)^i\Omega_0\exp[-(t-t_i)^2/\delta_t^2],
\end{equation}
where the relevant parameters are chosen as $t_1-t_2=6/\gamma_q$ and $\delta_t=5/\gamma_q$. The detuning is chosen as $\Delta_q=100\gamma_q$. The resulting Rabi frequencies experienced by each atom have the same Gaussian pulse shapes but with a different peak value, which we denote by $\Omega_{q}^{(s)}=\abs{e^{-\eta^2/2}\sin(\phi_s)\Omega_0}$ in accordance with Eq.~(\ref{eq:rabi}). A very useful property of STIRAP is that as long as $\Omega_q^{(s)}$ exceeds a certain threshold the state transfer is performed with a nearly ideal fidelity without introducing an unwanted phase change. In order to see this property, the fidelity of state transfer $\ket{\Psi}_s=\frac{1}{\sqrt2}(\ket{a}_s+\ket{b}_s)\rightarrow\ket{\Psi'}_s=\frac{1}{\sqrt2}(\ket{a}_s+\ket{q}_s)$ by STIRAP has been calculated numerically with varying $\Omega_q^{(s)}$. In Fig.~\ref{fig:stirap}, the fidelities ${}_s\!\bra{\Psi}\rho_s\ket{\Psi}_s$ (solid curve) and ${}_s\!\bra{\Psi'}\rho_s\ket{\Psi'}_s$ (dotted curve) are plotted, where $\rho_s$ is the final state after STIRAP. This figure clearly shows the desired result that the state $\ket{\Psi}_s$ is not altered when $\Omega_q^{(s)}$ is small whereas it is transferred to $\ket{\Psi'}_s$ when $\Omega_q^{(s)}$ is large enough. Consequently, the quenching operation $Q_L^{(k)}$ can be performed with a very high fidelity if we take $\Omega_0$ so that $\Omega_q^{(k+1)}$ is larger than the threshold (say, $20\gamma_q$ in Fig.~\ref{fig:stirap}). Note that here we do not have to worry about the upper bound of $\Omega_q^{(s)}$ since for large $\Omega_q^{(s)}$ the dynamics is dominated by $\Omega_{q1}^{(s)}$ and $\Omega_{q2}^{(s)}$, in which STIRAP depends only on the ratio between them, not on the individual amplitudes of them.

Thanks to the property of STIRAP, the fidelity of the quenching operation is not sensitive to the exact shapes or amplitudes of the applied standing-wave pulses. However, the fidelity would be affected more significantly by imprecise adjustments of the tilting angle $\theta_{qi}$ ($i=1,2$) and the relative phase between two plane waves. Let us denote the corresponding precisions respectively by $\Delta\theta$ and $\Delta\phi$. Suppose that we require every lattice sites in $S_L^{(k)}$ to coincide with the nodes within a spatial precision of $\Delta d$. Given a desired fidelity, the required precision $\Delta d$ can be determined from Fig.~\ref{fig:stirap}. For example, in order for the fidelity at each lattice site $s\in S_L^{(k)}$ to exceed the value corresponding to $\Omega_q^{(s)}=\gamma_q$ in Fig.~\ref{fig:stirap} with choosing $\Omega_q^{(k+1)}=20\gamma_q$,  the required precision is given by 
\vspace{-1mm}
\begin{equation}
\frac{\Delta d}{d_l}\lesssim\frac{1}{20}\mathrm{sinc}\left(\frac{\pi}{L+1}\right),
\vspace{-1mm}
\end{equation}
where $\mathrm{sinc}(x)\equiv\frac{\sin x}{x}$. If the total number of lattice sites is $N$, this precision is achieved when
\vspace{-1mm}
\begin{equation}
N\tan\theta_{qi}\,\Delta\theta+(L+1)\frac{\Delta\phi}{\pi} \lesssim \frac{\Delta d}{d_l}.
\vspace{-1mm}
\end{equation}
These inequalities give a trade-off relationship: in order to address a single atom in a larger optical lattice (larger $N$) with a less focused beam (larger $L$ and $\theta_{qi}$), we need more precise control of the tilting angle ($\Delta\theta$) and the relative phase ($\Delta\phi$) of the laser beams. For large $L$ and $\lambda_{qi}/2\sim d_l$, we reach an asymptotic condition
\vspace{-1mm}
\begin{equation}
(L+1)(N\Delta\theta+\Delta\phi/\pi)\lesssim1/20.
\vspace{-1mm}
\end{equation}
For large $N$, this condition imposes relatively high precision on the tilting angle. For example, if we take $L=4$ and $N=10^3$, we require $\Delta\theta\lesssim10~\mu\text{rad}$. This can be achieved through active stabilization based on a piezoelectic controller.

Finally, several remarks are in order. (a) In case of two dimensional optical lattices, the present scheme can be applied in the same manner by introducing more standing-wave fields assigned to each trap axis. (b) The quenching operation starts from an assumption that there is initially no population in state $\ket{q}$. After one round of operation $Q_L^{(k)}$, manipulation, and $(Q_L^{(k)})^{-1}$, however, some population unavoidably remains in state $\ket{q}$ due to the imperfections. In order to prevent this unwanted population from accumulating, it has to be optically pumped into the qubit subspace by using a globally addressing field after each round of operation (or possibly after every $n$ rounds for a moderate number $n$). Note that the optical pumping does not decrease the fidelity of the preceding operation. (c) For an arbitrary single-qubit rotation, three pairs of $Q_L^{(k)}$ and $(Q_L^{(k)})^{-1}$ are needed in general. It is because we use atoms having one quenching state as in Fig.~\ref{fig:level}. If we introduce another quenching state $\ket{q'}$ and perform the quenching operation such that $Q_L^{(k)}=\ket{q'}_s\!\bra{a}+\ket{q}_s\!\bra{b}$ for $s\not\in S_L^{(k)}$ \cite{vws06}, an arbitrary single-qubit rotation can be performed in one round of operation $Q_L^{(k)}$, manipulation, and $(Q_L^{(k)})^{-1}$. (d) A disadvantage of the present scheme is that all atoms have to be involved in addressing only one atom. Fortunately, the standard model of fault-tolerant quantum computation (FTQC) allows for such a situation since it regards both unitary operations and memories as being affected by the same amount of noise \cite{ste03}. Together with global addressing, the present scheme is well suited for performing collective operations of the form $\bigotimes_{s\in S_L^{(k)}}U^{(s)}$ and $\bigotimes_{s\in S_L^{(k)}}V^{(s,s+1)}$, where $U^{(s)}$ and $V^{(s,s+1)}$ are one- and two-qubit operations acting on lattice sites $s$ and $s+1$. If we take $L+1$ as the number of qubits in one error-correcting block, such collective operations are actually what should be done in the error-correcting steps of FTQC. In this context, further investigations on the effects of reduced parallelism and restricted geometry would be desirable.

\begin{acknowledgments}
The author thanks Hee Su Park, Sun Kyung Lee, and Sang-Kyung Choi for helpful discussions. This research was supported by the "Single Quantum-Based Metrology in Nanoscale" project of the Korea Research Institute of Standards and Science.
\end{acknowledgments}


\begin{references}

\bibitem{jz05} D. Jaksch and P. Zoller, Ann. Phys. \textbf{315}, 52 (2005).

\bibitem{jbc98} D. Jaksch, C. Bruder, J. I. Cirac, C. W. Gardiner, and P. Zoller, Phys. Rev. Lett. \textbf{81}, 3108 (1998).

\bibitem{gme02} G. Greiner \textit{et al.}, Nature \textbf{415}, 39 (2002).

\bibitem{sms04} T. St{\"o}ferle,  H. Moritz, C. Schori, M. K{\"o}hl, and T. Esslinger, Phys. Rev. Lett. \textbf{92}, 130403 (2004).

\bibitem{spp07} I. B. Spielman, W. D. Phillips, and J. V. Porto, Phys. Rev. Lett. \textbf{98}, 080404 (2007).

\bibitem{jbc99} D. Jaksch, H.-J. Briegel, J. I. Cirac, C. W. Gardiner, and P. Zoller, Phys. Rev. Lett. \textbf{82}, 1975 (1999).

\bibitem{bcj99} G. K. Brennen, C. M. Caves, P. S. Jessen, and I. H. Deutsch, Phys. Rev. Lett. \textbf{82}, 1060 (1999).

\bibitem{mgw03} O. Mandel \textit{et al.}, Nature \textbf{425}, 937 (2003).

\bibitem{rb01} R. Raussendorf and H. J. Briegel, Phys. Rev. Lett. \textbf{86}, 5188 (2001).

\bibitem{sch00} R. Scheunemann, F. S. Cataliotti, T. W. H{\"a}nsch, and M Weitz, Phys. Rev. A \textbf{62}, 051801(R) (2000).

\bibitem{dvm02} R. Dumke \textit{et al.}, Phys. Rev. Lett. \textbf{89}, 097903 (2002).

\bibitem{ppl03} S. Peil \textit{et al.}, Phys. Rev. A \textbf{67}, 051603(R) (2003).

\bibitem{mad06} Y. Miroshnychenko \textit{et al.}, Nature \textbf{442}, 151 (2006).

\bibitem{sdk04} D. Schrader \textit{et al.}, Phys. Rev. Lett. \textbf{93}, 150501 (2004).

\bibitem{zrs06} C. Zhang, S. L. Rolston, and S. Das Sarma, Phys. Rev. A \textbf{74}, 042316 (2006).

\bibitem{yc00} L. You and M. S. Chapman, Phys. Rev. A \textbf{62}, 052302 (2000).

\bibitem{cdj04} T. Calarco, U. Dorner, P. S. Julienne, C. J. Williams, and P. Zoller, Phys. Rev. A \textbf{70}, 012306 (2004).

\bibitem{jlb06} J. Joo, Y. L. Lim, A. Beige, and P. L. Knight, Phys. Rev. A \textbf{74}, 042344 (2006).

\bibitem{mlk97} C. Monroe \textit{et al.}, Phys. Rev. A \textbf{55}, R2489 (1997).

\bibitem{kts07} P. Kr{\'a}l, I. Thanopulos, and M. Shapiro, Rev. Mod. Phys. \textbf{79}, 53 (2007).

\bibitem{vws06} J. Volz \textit{et al.}, Phys. Rev. Lett. \textbf{96}, 030404 (2006).

\bibitem{ste03} A. M. Steane, Phys. Rev. A \textbf{68}, 042322 (2003).

\bibitem{jtd98} K. S. Johnson \textit{et al.}, Science \textbf{280}, 1583 (1998).

\end{references}
\end{document}